\documentclass[final,3p,times]{elsarticle}
\usepackage{graphicx}
\usepackage{amssymb}
\usepackage{amsmath}
\usepackage{lineno,hyperref}
\usepackage{color}
\hypersetup{colorlinks=true,linkcolor=red,citecolor=cyan}
\modulolinenumbers[200]

\journal{Journal Name}
\begin{document}
\begin{frontmatter}

\title{Rotating Curzon-Chazy metric}

\author[1,2]{Bobur Turimov\corref{cor}}\cortext[cor]{Corresponding author}\ead{bturimov@astrin.uz}
\author[1]{Odil Yunusov}\ead{odilbekhamroev@gmail.com}
\author[3]{Shavkat Karshiboev}\ead{shavkat.qarshiboyev.89@bk.ru}
\author[1,4,5]{Ahmadjon Abdujabbarov}\ead{ahmadjon@astrin.uz}
\address[1]{Ulugh Beg Astronomical Institute, Astronomy Str. 33, Tashkent 100052, Uzbekistan}
\address[2]{University of Tashkent for Applied Sciences, Str. Gavhar 1, Tashkent 100149, Uzbekistan}
\address[3]{Uzbek - Finnish pedagogical institute, Spitamen Shokh St. 166, Samarkand 140100, Uzbekistan}
\address[4]{Shahrisabz State Pedagogical Institute, Shahrisabz Str. 10, Shahrisabz 181301, Uzbekistan}
\address[5]{Tashkent State Technical University, Tashkent 100095, Uzbekistan}
\begin{abstract}
We demonstrate that the Curzon metric for a positive mass configuration possesses a singular event horizon with infinite area. This singularity has significant implications, revealing that the three-dimensional spatial hypersurfaces, which are orthogonal to the Killing vector field, exhibit a multiply connected structure. Furthermore, we investigate the dynamics of a test particle orbiting a central $\gamma$-object within this spacetime. It is found that under certain conditions, the particle's velocity can approach the speed of light, leading to an exceptionally high total energy at a specific value of the deformation parameter governing the spacetime structure. Moreover, we uncover a causality issue for a critical value of the deformation parameter, where the test particle can exceed the speed of light, potentially offering new insights into the theoretical existence of tachyons. This study contributes to the understanding of relativistic objects in deformed spacetimes and suggests that such violations of causality could play a role in explaining the elusive nature of tachyonic phenomena in high-energy physics.   
\end{abstract}

\end{frontmatter}

\date{\today}

\section{Introduction}

In general relativity, the Kerr metric represents one of the most important exact solutions of Einstein’s field equations, describing the geometry of spacetime around a rotating, uncharged black hole. Unlike the Schwarzschild metric, which models a non-rotating black hole with a spherically symmetric event horizon, the Kerr solution incorporates rotation, leading to a richer and more complex structure characterized by an ergosphere, frame-dragging effects, and the possibility of energy extraction via the Penrose process. The difficulty of obtaining exact analytical solutions for rotating compact objects stems from the nonlinear nature of Einstein's field equations. In many alternative theories of gravity, including modified gravity models and higher-dimensional theories, the inclusion of rotation often necessitates additional assumptions or approximations due to the increased complexity of the governing equations. Despite these challenges, the Kerr metric remains a fundamental solution in general relativity and serves as a foundation for exploring more general rotating solutions in modified theories. Extensions of the Kerr solution, such as those incorporating electric charge Kerr-Newman metric, higher dimensions Myers-Perry solution \cite{Myers:1986un}, or deviations due to modified gravity, are of great interest in theoretical physics and astrophysics. Studying these metrics provides insights into the nature of astrophysical black holes, the behavior of accretion disks, and the potential observational signatures of deviations from Einstein’s theory.

%%%%%%%%%%%%%%%%%%%%%%%%%%%%%%%%%%%%%%%%%%%%%%%%%%%%%%%%%%%%%%%%%%%%%%%%%%%%%%

The Curzon–Chazy spacetime, originally formulated as an exact solution to Einstein’s field equations, continues to be an intriguing subject of study, particularly in the context of alternative theories of gravity and its extensions. The introduction of a phantom scalar field counterpart to the Curzon–Chazy metric, as discussed in \cite{Polcar:2021}, further enriches the landscape of solutions by incorporating exotic matter fields, which may have implications for violations of the energy conditions and potential connections to wormhole physics or other exotic gravitational structures. An interesting characteristic of certain modifications to the Schwarzschild metric, such as the one presented in \cite{Janis:1968}, is the emergence of a singularity that deviates from the standard Schwarzschild horizon structure. In this case, the event horizon collapses into a singular point rather than forming a finite surface, leading to novel causal and geodesic structures. Such modifications provide insight into the nature of singularities in general relativity and the possible existence of naked singularities, which challenge the cosmic censorship conjecture. 

The study of geodesics and spatial curves in the Curzon metric has been explored in detail in \cite{Scott:1985}, where new properties of trajectories in this spacetime have been derived. Additionally, \cite{Morgan:1973} provides an in-depth examination of the behavior of geodesics near the singularity of the Curzon solution, shedding light on how test particles and light rays behave in this highly curved regime. Understanding the geodesic motion is crucial for determining possible observational consequences of such spacetimes, as well as for their stability and astrophysical relevance. In the presence of electromagnetic fields, generalizations of the Curzon–Chazy solution arise within the framework of the Einstein-Maxwell equations, as presented in \cite{Quevedo:PhysRevD}. These solutions introduce charged counterparts to the classical Curzon metric, which could be relevant in modeling astrophysical objects with strong electromagnetic fields, such as magnetars or charged compact objects in extreme conditions.

One of the most significant extensions of the Curzon-Chazy solution is the introduction of rotation, which has been investigated in \cite{Montero-Camacho:2014}. The rotating version of this metric is particularly relevant in understanding how frame-dragging and rotational effects modify the structure of singularities and geodesics in these spacetimes. Such rotating metrics could serve as approximations for the exterior gravitational field of rotating compact objects, providing potential applications in relativistic astrophysics and high-energy physics. These various extensions and analyses of the Curzon-Chazy spacetime contribute to a deeper understanding of alternative black hole geometries, singularity structures, and the role of additional fields such as scalar and electromagnetic fields in modifying classical solutions of general relativity.

%%%%%%%%%%%%%%%%%%%%%%%%%%%%%%%%%%%%%%%%%%%%%%%%%%%%%%%%%%%%%%%%%%%%%%%%%%%%%%

The study of static axially symmetric spacetimes within the framework of teleparallel gravity has provided alternative perspectives on gravitational energy-momentum definitions and the nature of curvature in alternative formulations of general relativity. In \cite{Sharif:2007}, the energy-momentum distribution for such spacetimes is discussed, highlighting how the teleparallel approach can provide insights into the physical interpretation of mass-energy in non-trivial spacetimes. A particularly interesting application of axially symmetric solutions is the construction of disk-like structures using the Miyamoto–Nagai and Chazy–Curzon potentials, incorporating a cut parameter to generate realistic astrophysical disk models, as examined in \cite{Coimbra-Araujo:2016}. Such models are highly relevant in studying the gravitational field of galaxies, accretion disks, and other astrophysical disk-like structures. Similarly, in \cite{Coimbra-Araujo:2007}, a Schwarzschild-Chazy-Curzon disk is extended into higher dimensions, providing a foundation for exploring how extra-dimensional theories may affect astrophysical disk solutions. 

The Simon-Mars scalars, which serve as important tools for identifying the presence of horizons and characterizing the deviations from Kerr-like spacetimes, have been used to analyze numerical solutions for boson stars and neutron stars in \cite{Some:2014}. These scalar quantities allow for the classification of numerical spacetimes and play a significant role in verifying whether a given solution corresponds to a black hole or an alternative compact object.

The existence of interior solutions matching the Curzon vacuum metric has been systematically studied in \cite{Bonnor:2013}, addressing a fundamental question about whether the Curzon solution, which traditionally describes an external gravitational field, can be extended to a physically meaningful interior region. The matching of interior and exterior solutions remains a crucial challenge in general relativity, particularly in the context of constructing viable models for compact astrophysical objects. In \cite{Gutierrez-Pinerez:2012f}, a simple model of active galactic nuclei (AGN) is analyzed, consisting of a black hole as the central engine. Given the relevance of supermassive black holes in AGN dynamics, the application of analytical spacetime solutions, including those derived from the Curzon-Chazy metric, can provide insights into the gravitational field structures that influence accretion processes and relativistic jet formation. In Ref. \cite{Abdelqader:2012} it has been shown that the entire field of the Curzon-Chazy solution, up to a region very close to the singularity, effectively mimics the Newtonian field of a ring. This finding reinforces the idea that certain relativistic metrics, despite their complex mathematical form, can have intuitive Newtonian analogue under specific conditions. This property is particularly useful for approximating relativistic effects in weak-field regimes while maintaining a connection to well-understood Newtonian gravity. These studies collectively highlight the significance of the Curzon-Chazy solution and its extensions in various contexts, ranging from theoretical investigations in modified gravity and extra dimensions to astrophysical applications involving galactic dynamics, compact objects, and high-energy astrophysics.

One of the simplest vacuum solutions of the Einstein field equations is the Curzon-Chazy metric described spacetime around a massive object of a mass $M$. In the spherical coordinates $x^\alpha=(t, r, \theta, \phi)$, the explicit form of the Curzon-Chazy metric is given as
\begin{align}\label{metric}
&ds^2=-\exp{\left(-\frac{2M}{r}\right)}dt^2+\exp{\left(\frac{2M}{r}\right)}\left[\exp{\left(-\frac{M^2\sin^2\theta}{r^2}\right)}(dr^2+r^2d\theta^2)+r^2\sin^2\theta d\phi^2\right]\ ,
\end{align}
which is a static metric and asymptotically flat as $r\to\infty$, with a Schwarschild mass $M$ and this metric belongs to Weyl’s class of solutions~\cite{Weyl1917AP,Weyl1919APa,Weyl1919APb}. It is worth noting that the structure of the Curzon-Chazy metric is very similar to the Papapetrou metric also known as the exponential metric \cite{Papapetrou1954ZP,Makukov2018PRD} that has the same form as in \eqref{metric} but without $\sin^2\theta$ in the exponential and in an equatorial plane these two metric are identical. However, the Papapetrou metric is the solution of Einstien-scalar fields equations and it is regular solution. In Ref.~\cite{Turimov2022PDU}, the radial dependence of the curvature scalars such Ricci, Kretchmann, and Weyl scalars in the Papapetrou metric has been explicitly discussed.  

One has to emphasise that the Curzon-Chazy metric is one of particular cases of the Zipoy-Voorhees solution is described by two parameters, $M$-mass of the object and $\delta$-deformation of spacetime. The Zipoy-Voorhees metric is given as~\cite{Zipoy1966JMP,Voorhees1970PRD}
\begin{align}\label{ZV}
ds^2&=-\left(1-\frac{2M}{\delta r}\right)^\delta dt^2+r^2\sin^2\theta\left(1-\frac{2M}{\delta r}\right)^{1-\delta}d\phi^2+\frac{(1-\frac{2M}{\delta r})^{\delta^2-\delta}}{(1-\frac{2M}{\delta r}+\frac{M^2\sin^2\theta}{\delta^2r^2})^{\delta^2-1}}\left(\frac{dr^2}{1-\frac{2M}{\delta r}}+r^2d\theta^2\right) ,
\end{align}
which is also known as $\delta$-metric ~\cite{Zipoy1966JMP,Voorhees1970PRD}, $\gamma$-metric~\cite{Shaikh2021,Chowdhury2012PRD,Abdikamalov2019PRD,Galtsov2019PRD,Herrera2000IJMPD,Toshmatov2019PRD,Benavides-Gallego2019PRD,Benavides-Gallego2020PRD}, and $q$-metric ~\cite{Toktarbay2014GC,Quevedo2015JMP,Boshkayev2016PRD,Frutos-Alfaro2018RSOS,Villamizar2021CQG}. In the following limiting case $\delta\to\infty$ the Zipoy-Voorhees metric reduces to the Curzon-Chazy metric. Notice that the Zipoy-Voorhees metric describes the naked singularity that is hypothetical compact object containing physical singularity uncovered with an event horizon. From the metric \eqref{ZV} one can find that the naked singularity is located at $r=2M/\delta$ and it tends to origin, i.e. $r\to 0$ as $\delta\to\infty$ (see e.g.\cite{Turimov2021Galax}). It concludes that the Curzon-Chazy metric does not contain the naked singularity.

The $\delta$-Kerr metric represents a modified version of the standard Kerr solution, incorporating deviations that may arise due to additional physical effects, alternative theories of gravity, or corrections to classical general relativity. These modifications alter the structure of spacetime surrounding a rotating black hole, potentially leading to new features in the event horizon, ergosphere, and the properties of geodesic motion. One of the primary motivations for studying the $\delta$-Kerr metric is to explore deviations from the classical Kerr solution, particularly in the strong-field regime. Such deviations may emerge from quantum gravity effects, extra-dimensional theories, or modifications to Einstein’s field equations. By analyzing the altered geometry and curvature, researchers can test general relativity’s predictions under extreme gravitational conditions and assess whether alternative models provide a better fit for observational data. From an astrophysical perspective, the $\delta$-Kerr metric offers a more flexible framework for modeling the dynamics of matter and radiation in the vicinity of black holes. The presence of additional parameters in the metric could influence accretion disk structure, relativistic jet formation, and gravitational wave signals from compact objects. Consequently, studying this modified metric is essential for interpreting observational data from modern telescopes and detectors, such as the Event Horizon Telescope and LIGO-Virgo collaborations.

Furthermore, the $\delta$-Kerr solution serves as a crucial tool for testing general relativity in the strong-field regime. If deviations from the standard Kerr metric are detected in black hole observations, they may provide evidence for new physics beyond Einstein’s theory. Such deviations could manifest in the form of changes in black hole shadow shapes, gravitational wave signatures, or modifications in the orbital dynamics of objects near the event horizon. Despite its theoretical significance, the $\delta$-Kerr metric remains an unconfirmed construct, requiring further analytical and numerical investigations to determine its viability and consistency with observational constraints. Ongoing research in this area aims to refine our understanding of gravity, test for potential deviations from general relativity, and explore the fundamental nature of spacetime around rotating black holes.

In the present paper, we are interested in deriving the rotating Curzon-Chazy metric using Kerr-$\delta$ metric. The paper is organized as follows. In Sect.~\ref{Kerrdelta}, we provide in very detailed derivation of the rotating Curzon-Chazy metric. In Sect.~\ref{Curzon}, we probe the rotating Curzon-Chazy metric. Finally, in Sect.~\ref{Summary}, we summarize obtained results. Throughout the paper, we use a space-like signature $(-,+,+,+)$ and a system of units in which $G=c=1$ (However, for those expressions with an astrophysical application we have written the speed of light explicitly.). Greek indices are taken to run from 0 to 3 and Latin indices from 1 to 3.

\section{Rotating Curzon-Chazy metric\label{Kerrdelta}}

Here we will focus one the deriving the rotating Curzon-Chazy spacetime using the $\delta$-Kerr metric which is the one of the vacuum solutions of Einstein field equations. It also belongs to the Weyl class of solutions and generalized form of the Kerr and Zipoy-Voorhees spacetimes. The explicit form of the $\delta$-Kerr metric given as~\cite{Allahyari2020CQG}
\begin{align}\label{deltaKerr}
ds^2=&-F(dt-\omega d\phi)^2+\frac{e^{2\gamma}B}{F}\left(\frac{dr^2}{A}+r^2d\theta^2\right)+\frac{A}{F}r^2\sin^2\theta d\phi^2\ ,
\end{align}
where unknown functions $A$, $B$, $F$, $\omega$ and $e^{2\gamma}$ are defined as 
\begin{align}\nonumber
&A=1-\frac{2M}{\delta r}+\frac{a^2}{\delta^2r^2}\ ,\qquad B=A-\frac{\sigma^2\sin^2\theta}{\delta^2 r^2}\ , \qquad F=\frac{\cal A}{\cal B}\ , \\\nonumber &\omega=2\left(a-\frac{\sigma}{\delta}\frac{\cal C}{\cal A}\right) \ ,\qquad e^{2\gamma}=\frac{1}{4}\left(1+\frac{M}{\sigma}\right)\frac{\cal A}{(x^2-1)^\delta}\left(\frac{x^2-1}{x^2-y^2}\right)^{\delta^2}\ , 
\end{align}
with
\begin{align}\nonumber
&{\cal A}=a_+a_-+b_+b_-\ ,\qquad {\cal B}=a_+^2+b_+^2\ ,
\qquad {\cal C}=(x-1)^q\left[x(1-y^2)(\lambda+\eta)a_++y(x^2-1)(1-\lambda\eta)b_+\right]\ ,
\\
&a_\pm=(x\pm 1)^q[x(1-\lambda\eta)\pm (1+\lambda\eta)]\ ,\qquad  b_\pm=(x\pm 1)^q[x(\lambda+\eta)\mp(\lambda-\eta)]\ , \qquad q=\delta-1\ ,
\\\nonumber
&\lambda=\alpha\frac{(x+y)^{2q}}{(x^2-1)^q}\ , \qquad \eta=\alpha\frac{(x-y)^{2q}}{(x^2-1)^q}\ , \qquad \alpha=\frac{a}{\sigma+M}=\frac{\sigma-M}{a}\ ,\qquad \sigma=\sqrt{M^2-a^2}\ ,
\end{align}
and the coordinates are defined as $x=(\delta r-M)/\sigma$ and $y=\cos\theta$. Notice that unlike other authors we have introduced the different constants of integration such as $M\to M/\delta$ and $\sigma\to\sigma/\delta$ that allows to get proper solution for rotating Curzon-Chazy metric. As we mentioned before that $\delta$-Kerr metric is generalized form of the Kerr and Zipoy-Voorhees spacetimes. By substituting $\delta=1$ one can get the Kerr spacetime, while in the case when $a=0$ the metric \eqref{deltaKerr} reduces to Zipoy-Voorhees spacetime which is given in \eqref{ZV}. Therefore $\delta$-Kerr metric is described not only black hole but also rotating naked singularity. Let's check the $\delta$-Kerr spacetime in the following limiting case $\delta\to\infty$ and produce the following results:
\begin{align}\label{RCC}\nonumber
&\lim_{\delta\to\infty} g_{tt}=-\exp{\left(-\frac{2\sigma}{r}\right)} \ ,\\\nonumber
&\lim_{\delta\to\infty} g_{t\phi}=2a\exp{\left(-\frac{2\sigma}{r}\right)} \ ,\\
&\lim_{\delta\to\infty} g_{rr}=\exp{\left(\frac{2\sigma}{r}-\frac{\sigma^2\sin^2\theta}{r^2}\right)}\ , \\\nonumber
&\lim_{\delta\to\infty} g_{\theta\theta}=\exp{\left(\frac{2\sigma}{r}-\frac{\sigma^2\sin^2\theta}{r^2}\right)}r^2\ , \\\nonumber
&\lim_{\delta\to\infty} g_{\phi\phi}=\exp{\left(\frac{2\sigma}{r}\right)}r^2\sin^2\theta-4a^2\exp{\left(-\frac{2\sigma}{r}\right)}\ .
\end{align}
To better understand the spacetime formation in \eqref{RCC}, one can analyse the Kretchmann scalar $K=R_{\alpha\beta\mu\nu}R^{\alpha\beta\mu\nu}$ which can be expressed as  
\begin{align}
K=&\frac{48\sigma^2}{r^6}\left[\left(1-\frac{\sigma}{r}\right)^2+\frac{\sigma^2\sin^2\theta}{r^2}\left(1-\frac{\sigma}{r}+\frac{\sigma^2}{3r^2}\right)\right]\exp{\left(-\frac{2\sigma}{r}+\frac{\sigma^2\sin^2\theta}{r^2}\right)} \ ,
\end{align}
and in the following limiting case when $r\to 0$, it is divergent that means the spacetime metric \eqref{RCC} is singular solution to Eisenstein field equations. Using the expressions for the components of the metric tensor in \eqref{RCC}, the simplified form of the rotating Curzon-Chazy metric is rewritten as 
\begin{align}\label{metric1}
&ds^2=-\exp{\left(-\frac{2\sigma}{r}\right)}(dt-2ad\phi)^2+\exp{\left(\frac{2\sigma}{r}\right)}\left[\exp{\left(-\frac{\sigma^2\sin^2\theta}{r^2}\right)}(dr^2+r^2d\theta^2)+r^2\sin^2\theta d\phi^2\right]\ ,
\end{align}
which is very similar to the static Curzon-Chazy spacetime given in \eqref{metric}. Indeed, one can easily check that after performing the following transformations $\sigma\to M$ and $t-2a\phi\to t$ in \eqref{metric1} reduces to the static Curzon-Chazy spacetime. In the case of maximal rotating object (i.e. $a\to M$ or $\sigma=0$) the solution takes a form: 
\begin{align}\label{metric2}
&ds^2=-dt^2+dr^2+r^2(d\theta^2+\sin^2\theta d\phi^2)\ ,
\end{align}
which is Minkowski spacetime.

\section{Geodesics motion\label{Curzon}}

Now let us examine the rotating Curzon-Chazy spacetime while considering the motion of a test particle. Taking into account the above fact, the formation of the Curzon-Chazy spacetime can be understood as equivalent to the following static spacetime:

\begin{align}\label{metric0}
ds^2 = -\exp{\left(-\frac{2\sigma}{r}\right)}dt^2 + \exp{\left(\frac{2\sigma}{r}\right)}\left[\exp{\left(-\frac{\sigma^2\sin^2\theta}{r^2}\right)}(dr^2 + r^2d\theta^2) + r^2\sin^2\theta d\phi^2\right]\ .
\end{align}
This metric is very similar to the Curzon-Chazy metric itself, with the only difference being that the mass parameter of the gravitational object is replaced by $\sigma$. Therefore, it is interesting to consider the geodesic motion in this spacetime and compare it with what has been previously obtained in the Schwarzschild spacetime. 

Now we focus on the geodesic motion in the rotating Curzon-Chazy spacetime spacetime. The conserved quantities, namely, the energy and angular momentum of test particle in this spacetime are given by
\begin{align}
&{\cal E}=\exp{\left(-\frac{2\sigma}{r}\right)}{\dot t}\ , \qquad {\cal L}=\exp{\left(\frac{2\sigma}{r}\right)}r^2\sin^2\theta{\dot\phi}\ .
\end{align}
Since the spacetime (\ref{metric2}) is spherically-symmetric, one can easily consider particle motion at the equatorial plane, i.e. $\theta=\pi/2$. The equation for the radial motion for a particle in spacetime (\ref{metric2}) is written as~\cite{Boonserm2018PRD}
\begin{align}\label{radial}
\exp{\left(\frac{\sigma^2\sin^2\theta}{r^2}\right)}{\dot r}^2={\cal E}^2-\exp{\left(-\frac{2\sigma}{r}\right)}\left[1+\frac{{\cal L}^2}{r^2}\exp{\left(-\frac{2\sigma}{r}\right)}\right]\ ,    
\end{align}
where ${\cal E}$, ${\cal L}$ are, respectively, the specific energy and specific angular momentum of test particle. The overdot denotes the derivative with respect to an affine parameter. 

\begin{figure}
    \centering
    \includegraphics[width=.4\textwidth]{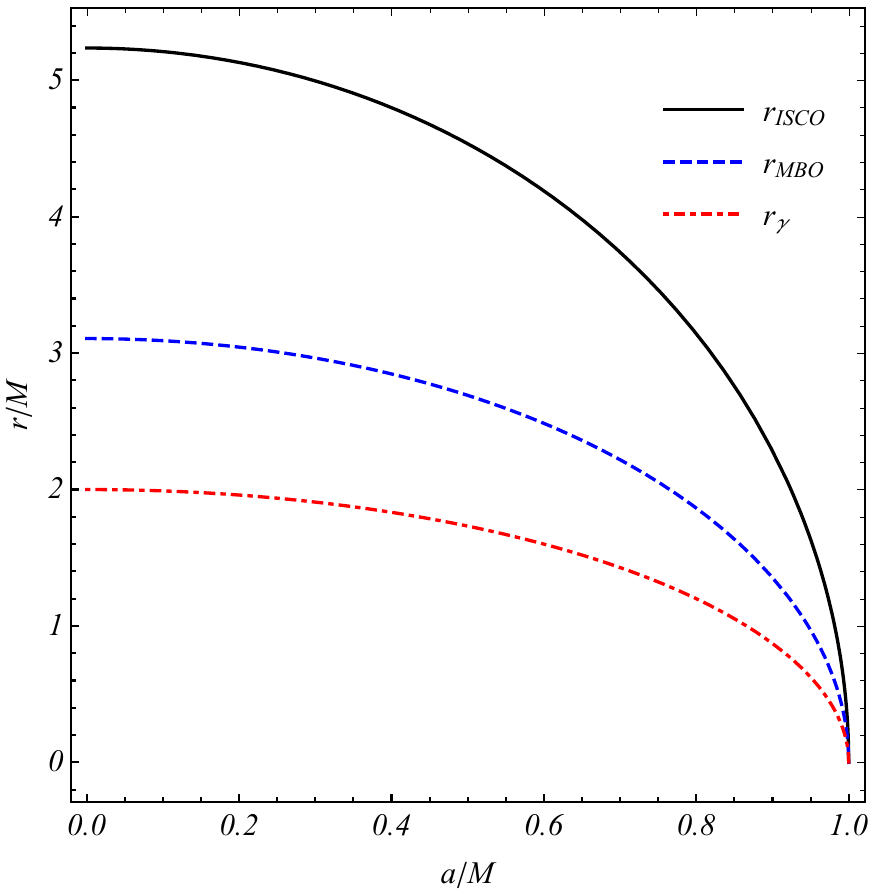}
    \includegraphics[width=.4\textwidth]{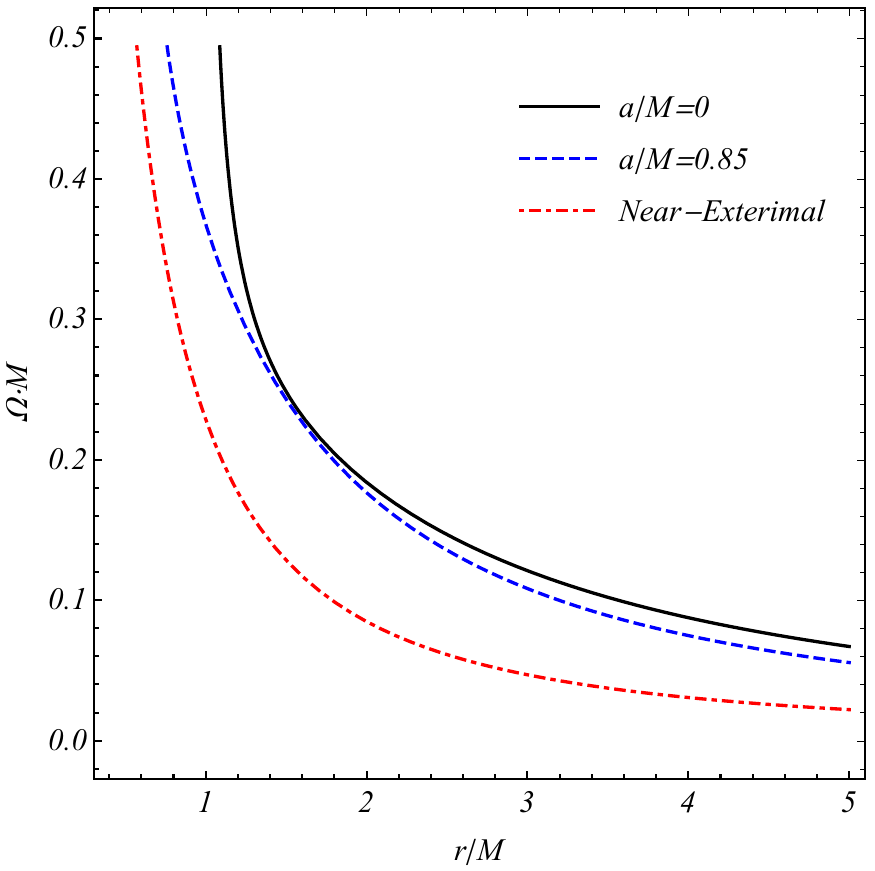}
    \caption{(Left panel) Dependence of the characteristic radii (ISCO, MBO, photon-sphere) with respect to spin parameter $a$ and (Right panel) dependence of Keplerian angular velocity with respect to radial distance for different values of spin  parameter $a$.}
    \label{omega}
\end{figure}
\begin{figure}
    \centering
    \includegraphics[scale=0.45]{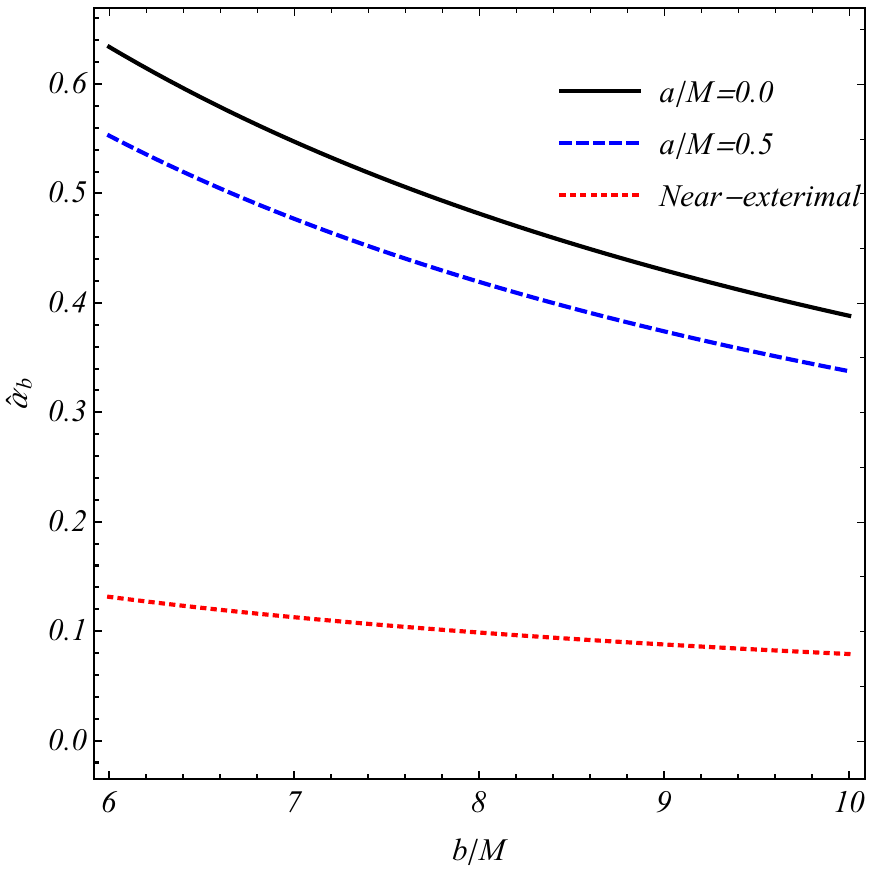}
    \includegraphics[scale=0.45]{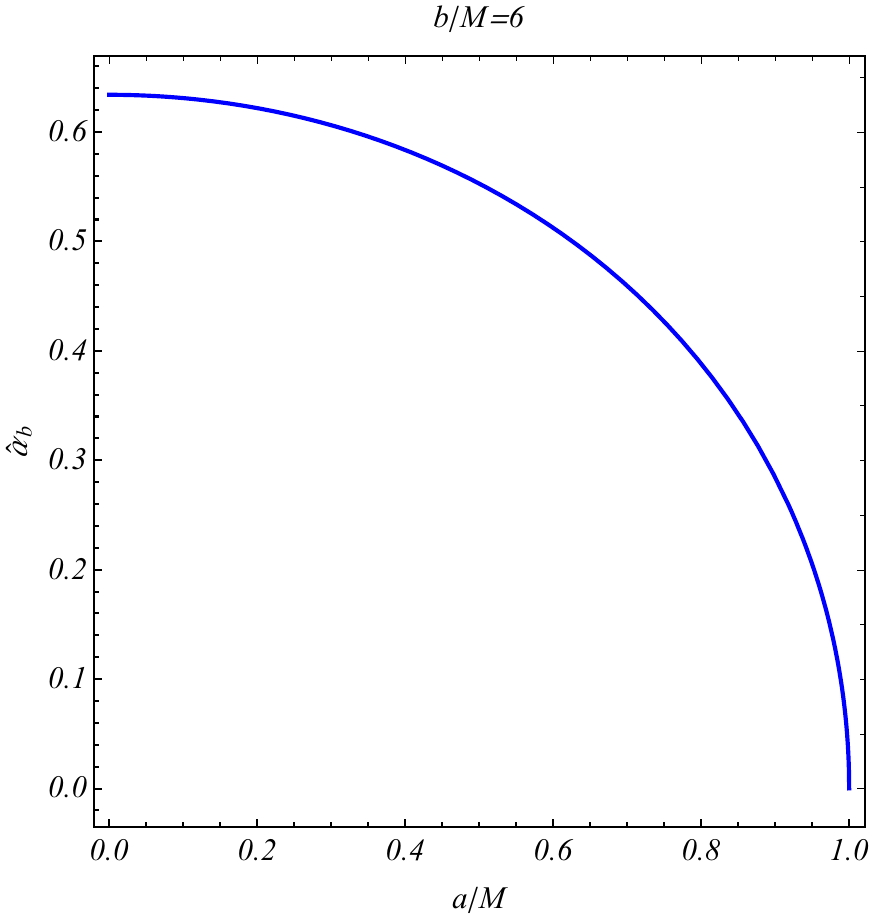}
    \caption{(Left panel) Plot of the deflection angle as a function of $b$ for different values of $a$ and (Right panel) $a$ with a fixed impact parameter.}\label{defl.angle}
\end{figure}

From the astrophysical point of view it is important to find the stable orbit of test particle around the gravitational compact object. In many astrophysical situations, it is widely believed that the position of the inner edge of the accretion disk around central object to be located at the innermost stable circular orbit (ISCO). From this point of view finding the ISCO position is very important task. Note that the following conditions ${\dot r}={\ddot r}=0$ are also valid for particles and the critical value of the specific energy and specific angular momentum then take form:
\begin{align}
{\cal E}^2(r)=\frac{r-\sigma}{r-2\sigma}e^{-2\sigma/r}\ ,\qquad {\cal L}^2(r)=\frac{\sigma r^2}{r-2\sigma}e^{2\sigma/r}\ ,    
\end{align}
and the ISCO radius for particle orbiting around cdentral object can be determined, from the conditions ${\cal E}'(r)={\cal L}'(r)=0$, as~\cite{Boonserm2018PRD}
\begin{align}\label{isco}
r_{\rm ISCO}^\pm=(3\pm\sqrt{5})\sigma\ .     
\end{align}
which is less than the ISCO position in the Schwarzschild space i.e. $r_{\rm ISCO}=6M$. Note that the physically meaningful solution for the ISCO position for test particle is $r_+\simeq 5.236\sigma$, because $r_-\simeq 0.764\sigma$ is less that the position of the photon-sphere in this spacetime. Interestingly, for the maximally rotating Curzon-Chazy spacetime, i.e, $a=M$ or $\sigma=0$, the ISCO position for test particle is to be zero. Because in this case the spacetime reduces to the Minkowskii spacetime. 

One has to emphasise that there is other type of orbit for test particle so-called the marginally bound orbit (MBO) which can be found from the condition, ${\cal E}(r)=1$, as the solution of the following nonlinear equation:
\begin{align}
e^{2\sigma/r}=\frac{r-\sigma}{r-2\sigma}\ ,\quad\to\quad r_{\rm MBO}\simeq 3.10657\sigma\ ,    
\end{align}
which is also less than that obtained in the Schwarzschild spacetime i.e. $r_{\rm MBO}=4M$.

In the present paper, it is assumed that test particle follows Keplerian orbit around the wormhole. One of the significant feature of the accretion disk around worhole is the energy efficiency so-called the gravitational defect mass calculated as, i.e., $\eta=1-{\cal E}_{\rm ISCO}$, or 
\begin{align}
\eta=1-\frac{\sqrt{5}+2}{\sqrt{5}+1}e^{-2/(3+\sqrt{5})}\simeq 10.7\%\ ,
\end{align}
while in the Schwarzschild space it is $\eta\simeq 5.72\%$. One of the observable quantity, so-called Keplerian frequency in the exponential spacetime can be determined as
\begin{align}
\Omega=\sqrt{\frac{\partial_r g_{tt}}{\partial_r g_{\phi\phi}}}=\frac{1}{r}\sqrt{\frac{\sigma}{r-\sigma}}e^{-2\sigma/r}\ ,    
\end{align}
while in the Schwarzschild space, it reduces to $\Omega=\sqrt{M/r^3}$. The left panel of Fig.~\ref{omega} illustrates the behavior of characteristic radii, including the innermost stable circular orbit (ISCO), the marginally bound orbit (MBO), and the photon sphere, as functions of the spin parameter $ a $. As observed, all these radii decrease monotonically with increasing $ a $ and eventually shrink to zero in the extremal Curzon-Chazy spacetime. This result is consistent with the fact that in the extremal limit, the Curzon-Chazy metric effectively reduces to Minkowski spacetime, where no characteristic radii exist, as there is no gravitational field to sustain bound orbits.  

In the right panel of Fig.~\ref{omega}, the radial dependence of the angular velocity of a massive test particle in the Curzon-Chazy spacetime is depicted for various values of the spin parameter. The plot reveals that as the spin parameter increases, the angular velocity of the orbiting particle decreases. In the extremal Curzon-Chazy spacetime, where the metric asymptotically transitions to Minkowski space, the angular velocity ultimately vanishes. This suggests that in this limiting case, the gravitational influence of the central object diminishes, leading to the absence of stable orbital motion. These findings highlight the distinct nature of the Curzon-Chazy spacetime compared to the standard Kerr solution. The disappearance of characteristic radii in the extremal case indicates a fundamental departure from the behavior observed in the Kerr geometry, where extremal solutions still support well-defined orbital structures. Such deviations could have significant implications for observational signatures of rotating compact objects and for testing alternative theories of gravity.

\section{Null geodesics}

We now consider photon motion, (i.e. $m=0$) in the rotating Curzon-Chazy spacetime (\ref{metric}). One can immediately find that the equation for radial motion in the equatorial plane reads
\begin{align}
\exp{\left(\frac{\sigma^2}{r^2}\right)}{\dot r}^2={\cal E}^2-\frac{{\cal L}^2}{r^2}\exp{\left(-\frac{4\sigma}{r}\right)}\ ,    
\end{align}

The stationary and turning points of photon are governed by the following conditions, ${\dot r}={\ddot r}=0$, that allows to find location of photon-sphere, $r_{\rm ph}$, and impact parameter, $b={\cal L}/{\cal E}$ in the form: $r_{\rm ph}=2\sigma$ and $b=r_{\rm ph}e^{2\sigma/r_{\rm ph}}=2e\sigma$. Note that in the Schwarzschild space these quantities take the form: $r_{\rm ph}=3M$ and $b=3\sqrt{3}M$. It is interesting to study the capture cross section of photon (area of shadow) by the wormhole, $\sigma_{cs}=\pi b^2$ which takes a form: 
\begin{align}
\sigma_{cs}=4\pi e^2(M^2-a^2)\simeq 92.8536(M^2-a^2)\ ,
\end{align}
and it is greater than cross section of photon by Schwarzschild black hole with identical mass, i.e. $\sigma_{\rm Sch.}=27\pi M^2\simeq 84.823M^2$. Notice that previous calculations are related to the strong gravitational lensing in the rotating Curzon-Chazy spacetime. It is also important to test this spacetime with a weak gravitational lensing. It is well-known that gravitational lensing is one of the powerful tools to test the Einstein theory of general relativity (GR) versus alternate theories of gravity. According to GR theory the deflection of the light-ray around static spherically symmetric gravitational source with the total mass $M$ is $\hat\alpha=4M/b$, where $b$ is the impact parameter of the light-ray. Here we study the weak gravitational lensing effect in Curzon-Chazy spacetime. The one of simple approaches for studying gravitational lensing has been developed in~\cite{Bisnovatyi-Kogan, Morozova13, Turimov2019IJMPD, Ahmedov2019IJMPS}, where the metric tensor can be expanded as $g_{\alpha\beta}\simeq\eta_{\alpha\beta}+h_{\alpha\beta}$ in the weak field approach, where $\eta_{\alpha\beta}$ is the metric tensor of a flat Minkowski space-time and $h_{\alpha\beta}$ is a perturbation of the metric tensor. Accordingly the deflection angle of the light-ray can be determined as~\cite{Bisnovatyi-Kogan}
\begin{align}
\hat\alpha_\alpha=\frac{1}{2}\int_{-\infty}^{\infty} \left(\frac{\partial h_{tt}}{\partial x^\alpha}+\frac{\partial h_{zz}}{\partial x^\alpha}\right)dz\ , 
\end{align}
where the metric perturbation in the rotating Curzon-Chazy is given by $h_{tt}=2\sigma/r$ and $h_{zz}=2\sigma/r-\sigma^2 \sin^2{\theta}/r^2$. We assume that the light-ray approaches to the compact object along z-direction. Hereafter introducing the impact parameter of the light-ray, i.e., $b$, the radial coordinate takes a form $r=\sqrt{z^2+b^2}$. Then the expression for the deflection angle takes a form:
\begin{align}\label{def}
\hat\alpha_b = -\int _{-\infty}^{+\infty}\frac{b
}{2r}\frac{d}{dr}\left(\frac{4\sigma}{r}-\frac{\sigma^2\sin^2{\theta}}{r^2}\right)dz=
\frac{4\sigma}{b}-\frac{3\pi}{8}\frac{\sigma^2}{b^2}\ , 
\end{align}
which is first term the same expression as obtained in general relativity, however, the sign should be opposite and the second term which is adition due to exponential term in the metric. Since the deflection angle is a vector quantity, its sign can be easily adjusted in the expression (\ref{def}). The left panel of Fig.~\ref{defl.angle} illustrates the dependence of the deflection angle on the impact parameter $ b $. As shown, the deflection angle decreases as $ b $ increases, indicating that light rays passing farther from the central object experience weaker gravitational bending, consistent with expectations from general relativity.  In the right panel of Fig.~\ref{defl.angle}, the effect of the rotation parameter $ a $ on the deflection angle is depicted. As the rotation parameter increases, the deflection angle gradually decreases, signifying a reduction in the gravitational lensing effect. In the limit of maximal rotation, corresponding to the extremal Curzon-Chazy spacetime, the deflection angle becomes zero. This implies that in this extreme case, light rays travel in straight paths as if the spacetime were Minkowskian, further reinforcing the idea that the extremal Curzon-Chazy solution effectively reduces to flat space. These results highlight significant deviations from the standard Kerr spacetime, where even in the extremal case, light still experiences gravitational bending due to the presence of a well-defined event horizon. The vanishing deflection angle in the extremal Curzon-Chazy case suggests that its gravitational influence is fundamentally different, which could have implications for gravitational lensing observations and tests of alternative spacetime geometries.
\begin{figure}
    \centering
    \includegraphics[width=.32\textwidth]{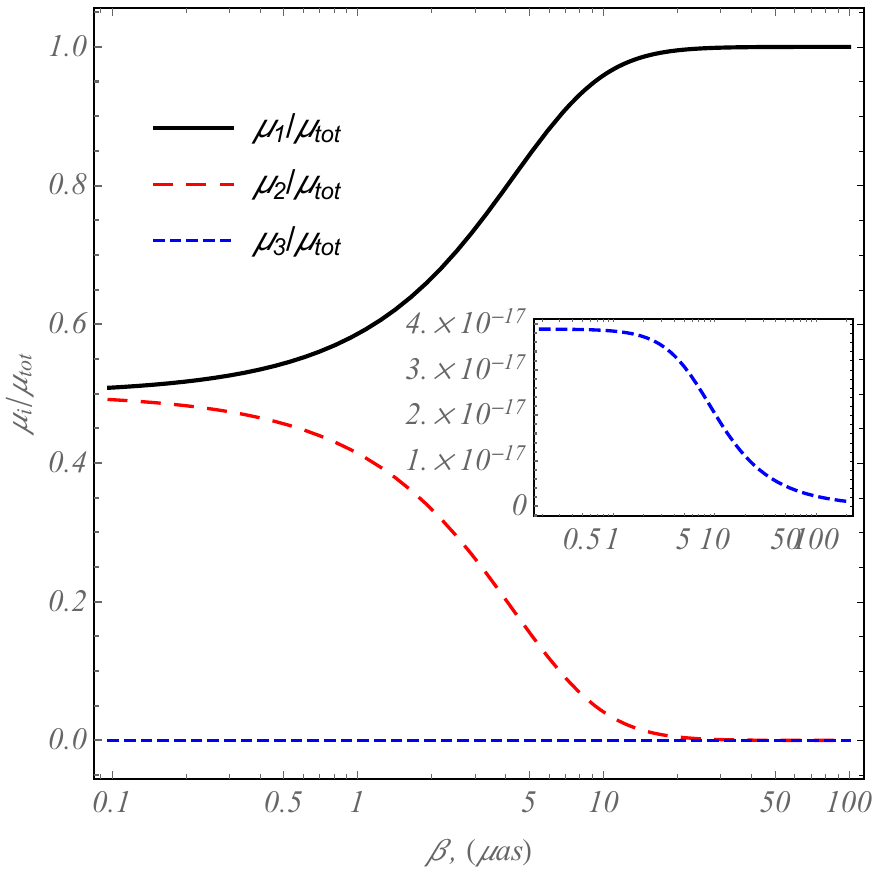}
    \includegraphics[width=.32\textwidth]{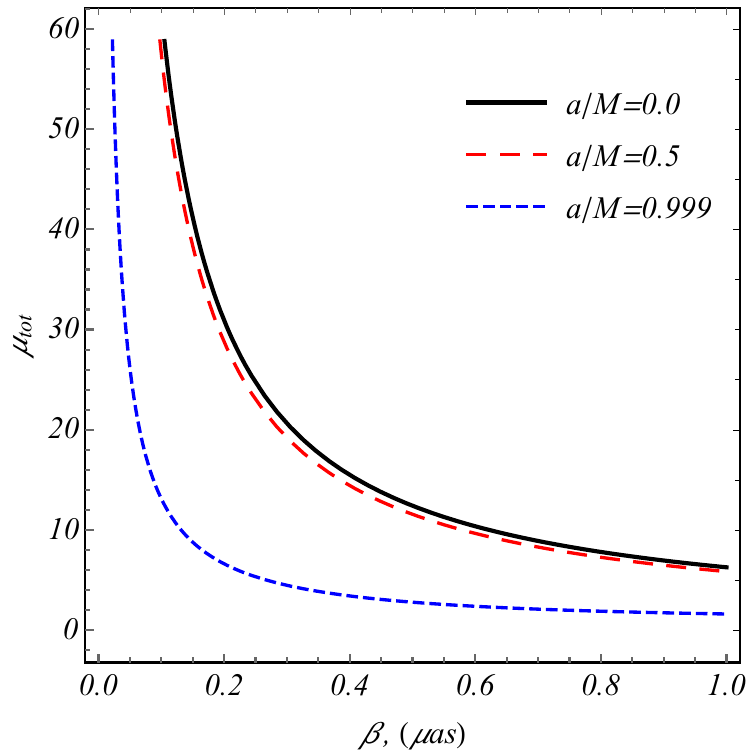}
    \includegraphics[width=.32\textwidth]{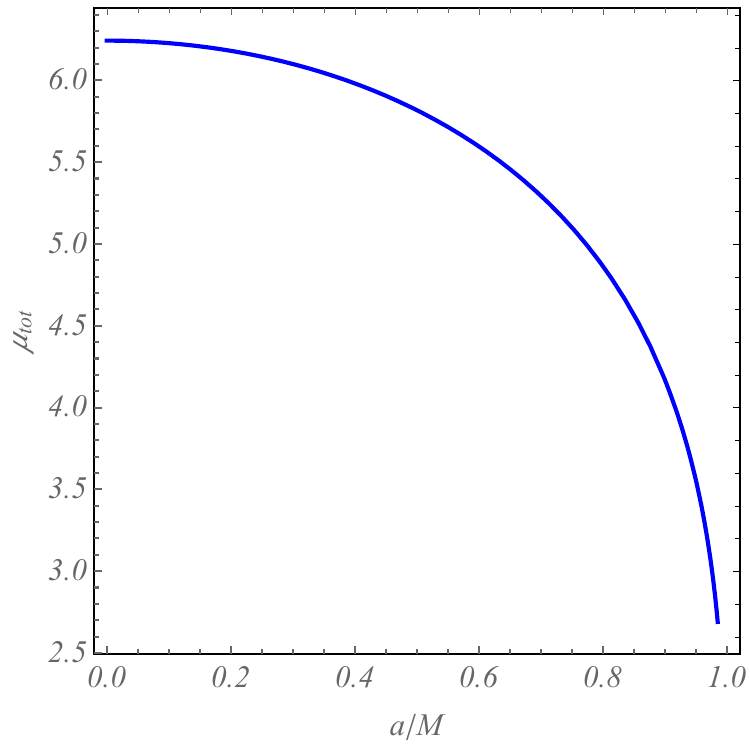}
    \caption{The dependence of the ratio of the main magnification to total one for the different values of parameter $a$. Blue lines are represent $\mu_{+}/\mu_{total}$ while red lines are show $\mu_{-}/\mu_{total}$. The magnifications of the source are presented for different spin parameter a of the gravitational object. The magnifications of the source are presented for different spin parameter a of the gravitational object. Where we have taken  $M/D_d=1.911\cdot 10^{-11}$ and $D_s/D_{ds}=2$.}
    \label{magnification}
\end{figure}

It is known that the gravitational lens system contains three fragments: the source of the light-ray, gravitational object and observer. If these fragments are located in the same line then one can observe Einstein ring, otherwise one can observe two images of the source. According to the standard model of the gravitational lensing, the angular half-separation due to gravitational lensing between the images of the source is proportional to the Einstein deflection angle (\ref{def}), and it can be expressed as~\cite{Schneider92}
\begin{align}
\Theta=\sqrt{\frac{4MD_{ds}}{D_sD_d}}\ ,    
\end{align}
where $D_d$ is the distance between observer and lens, $D_s$ is the distance between observer and source, and $D_{ds}$ is the distance between the lens and source.

In the context of the Curzon-Chazy spacetime, the lensing effect is significantly altered compared to standard general relativistic solutions, such as the Schwarzschild or Kerr metrics. As previously discussed, the deflection angle decreases with increasing rotation parameter $a$ and eventually vanishes in the extremal Curzon-Chazy case. This suggests that for a maximally rotating Curzon-Chazy object, the gravitational lensing effect disappears entirely, leading to the absence of multiple images or an Einstein ring. This deviation from the predictions of the Kerr metric has potential observational consequences. If a rotating compact object were to exhibit such lensing behavior, it could provide a way to distinguish between different spacetime geometries and test the validity of alternative gravitational models. Future gravitational lensing surveys, such as those conducted by the Vera C. Rubin Observatory or space-based missions, could potentially detect such anomalies, offering new insights into the nature of strong gravitational fields.
\begin{figure}
\centering
\includegraphics[width=.4\textwidth]{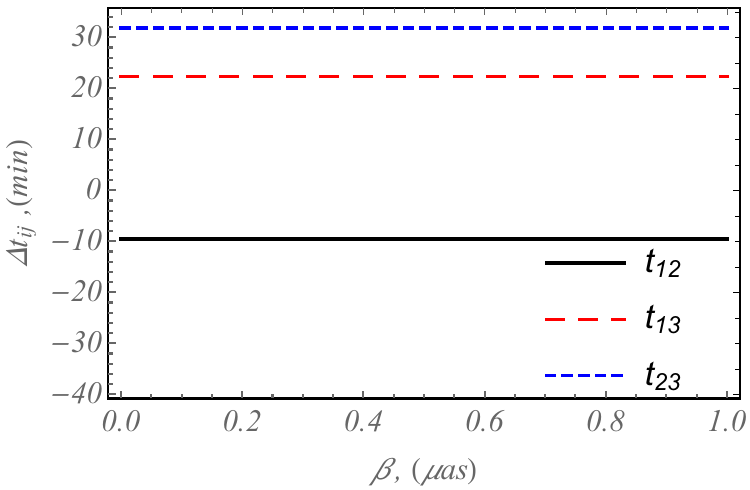}
\caption{Dependence of the time delay due to gravitational field from the  position $y$ for $z=1.7$ Where we have used $M/D_d=1.911\cdot 10^{-11}$ and $\beta=1 \,(\mu as)$.\label{TimeDelay}}
\end{figure}
\begin{figure}
  \centering
    \includegraphics[width=.475\textwidth]{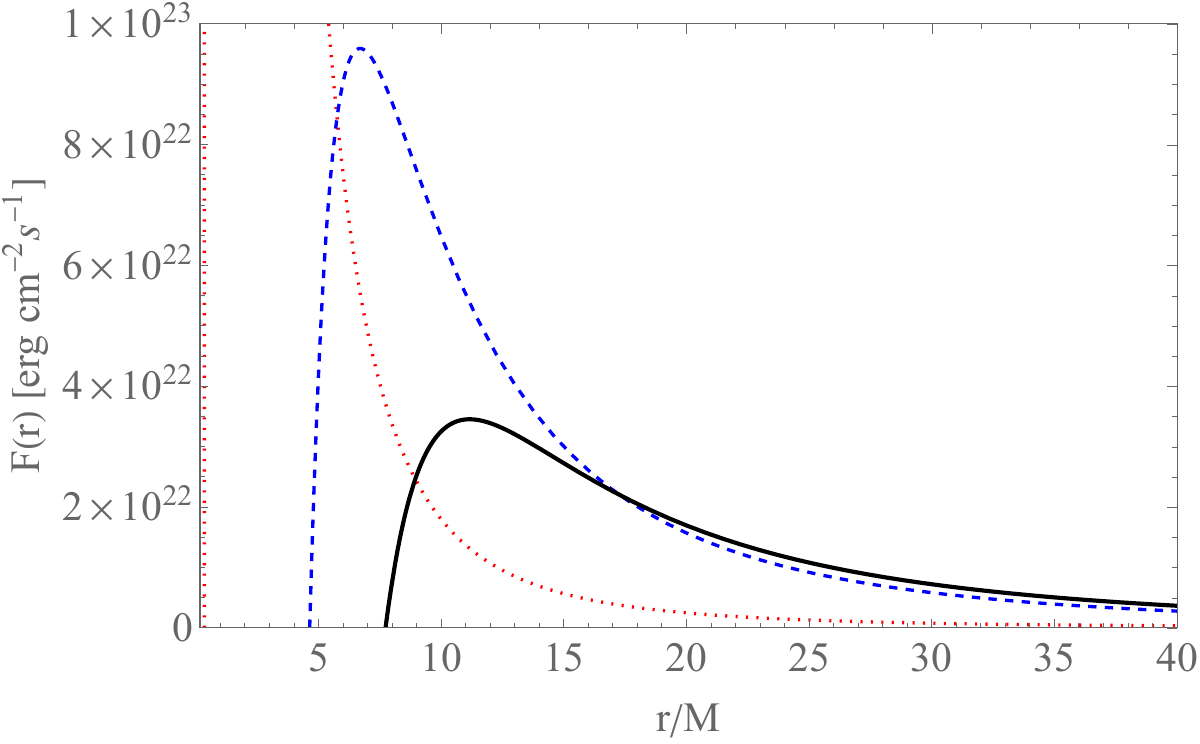}
    \includegraphics[width=.45\textwidth]{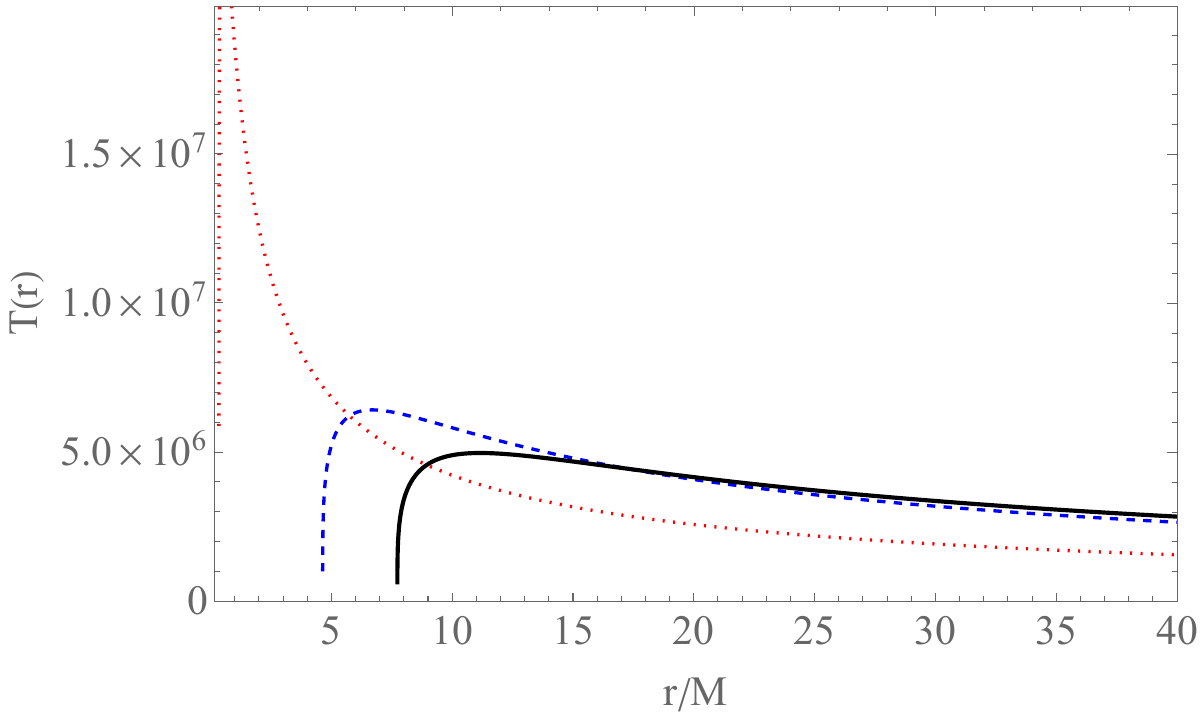}
    \caption{The energy flux $F(r)$ , Upper panel and temperature $T(r)$, lower  panel of a disk around black hole characterized by rotating Curzon metric for different values of a. Black thick, Blue dotdashed and red dotted lines describe the case of $a=0$, $a=0.8$, $a=0.998$, respectively. }
    \label{energy flux}
\end{figure}

It is extremely interesting to study the observational consequences of the gravitational lensing, namely, the magnification of image sources. For this purpose, one can consider the lens equation as in below form: 
\begin{align}
\beta=\theta-\frac{D_{ds}}{D_s}\alpha(\theta)\ ,
\end{align} 
where the angle $\beta$ is the real object from the observer-lens axis and angle $\theta$ is the apparent image of the object from the observer-lens axis, respectively. In the weak lensing approach, we assume that $b=\theta D_{s}$. Using eqs.(\ref{def}), we can rewrite equation (21) that form:
\begin{align}
\theta^3-\beta\theta^2-\theta_{0}^2\theta-\Phi=0\ ,
\end{align}
where $\theta_{0}^2=\sigma\Theta^2/M$ and $\Phi=3\pi\sigma^2\Theta^2/(8MD_d)$. One can reform this equation as follows $x^3+px+q=0$, where $x=\theta-\beta/3$\ , $p=-\theta_{0}^2-\beta^2/3$, and $q=-\theta_{0}^2 \beta/3-2\beta^3/27-\Phi$. One can seek solution to the cubic equation takes a form: 
\begin{align}
x_k=2s^{1/3}\cos{\frac{\gamma+2\pi k}{3}}\ ,\qquad s=\sqrt{-\frac{p^3}{27}}\ ,\qquad \cos{\gamma}=-\frac{q}{2s}\ , \qquad k=0, 1, 2\ .
\end{align}
Finally, the total magnification of the gravitationally lensed images is determined using the formula (\text{See e.g.}~\cite{Schneider92}):

\[
\mu = \sum_{k=0}^{3} \left| \frac{\theta_k}{\beta} \frac{d\theta_k}{d\beta} \right| \quad ,
\]
where $ \theta_k $ represents the angular positions of the images and $ \beta $ is the angular separation between the source and the lens. In Fig.~\ref{magnification}, the dependence of the ratio of individual image magnifications to the total magnification is shown. It is evident from the figure that as the angular separation $ \beta $ increases, the magnification of the main image becomes dominant. For large values of $ \beta $, the main image magnification approaches the total magnification, as the contributions from the other images diminish. Additionally, Fig.~\ref{magnification} demonstrates the relationship between magnification and the radial position $ y $. It also highlights that the total magnification $ \mu $ decreases as the spin parameter $ a $ increases. This trend indicates that higher spin values lead to a reduced overall magnification, which can be attributed to the altered gravitational lensing properties in rotating spacetimes. In the case of a maximally rotating Curzon-Chazy object, the magnification reaches its minimum value, further emphasizing the unique characteristics of this spacetime solution. This reduction in magnification with increasing spin could potentially be detected in gravitational lensing observations, offering another avenue for distinguishing different black hole models and gaining insights into the nature of the spacetime surrounding compact objects.

On the other hand, the geometrical time delay is determined as~\cite{Schneider92}
\begin{align}
&\Delta t_{12}=-\sqrt{3}\frac{2M}{\Theta^2}(1+z)\sin{\frac{\gamma+\pi}{3}}\left(\cos{\frac{\gamma+\pi}{3}}-\frac{4\beta}{3}\right)\ ,\\
&\Delta t_{13}=\sqrt{3}\frac{2M}{\Theta^2}(1+z)\sin{\frac{\gamma+\pi}{3}}\left(\cos{\frac{\gamma+2\pi}{3}}+\frac{4\beta}{3}\right)\ ,\\
&\Delta t_{12}=-\sqrt{3}\frac{2M}{\Theta^2}(1+z)\sin{\frac{\gamma}{3}}\left(\cos{\frac{\gamma}{3}}+\frac{4\beta}{3}\right)\ .    
\end{align}
where $z$ is the red-shift factor. The dependence of the geometrical time delay on the position $ y $ for a redshift value of $ z = 1.7 $ is illustrated in Fig.~\ref{TimeDelay}. As evident from the plot, the time delay between the first and second images, as well as between the first and third images, remains equal. Additionally, as the lens-source angular misalignment $ \beta $ increases, the overall time delay also increases.  An important feature of the time delay behavior is its dependence on the spin parameter $ a $. As the spin increases, particularly when $ a $ approaches its maximum value, the time delay grows at a faster rate compared to the non-rotating case ($ a = 0 $). This suggests that the presence of rotation significantly affects the gravitational potential, leading to a more pronounced delay in photon trajectories. In the context of gravitational lensing, time delay measurements provi de crucial insights into the underlying spacetime geometry. The observed increase in time delay with rotation could serve as an important observational signature distinguishing the Curzon-Chazy spacetime from other models, such as the Kerr metric. Future precise measurements of time delays in strong lensing systems, particularly with high-redshift quasars or fast radio bursts, could provide further constraints on the rotation parameter and help test alternative theories of gravity.

\section{Thin accretion disk\label{sec:Results}}

A thin accretion disk is a theoretical model used to describe the structure and behavior of matter falling onto a compact object, such as a black hole or a neutron star, through gravitational attraction. The term "thin" refers to the assumption that the disk has a small aspect ratio, meaning that its thickness is much smaller than its radial extent. In a thin accretion disk, the infalling matter forms a flat, rotating disk-like structure around the central object. The disk is composed of gas, plasma, and dust particles that gradually lose angular momentum and spiral inward due to energy dissipation processes, such as viscosity and magnetic torques. The dynamics of a thin accretion disk are governed by several physical processes. It allows matter to move inward while releasing gravitational potential energy, which is then converted into thermal radiation. The disk's temperature profile is another important factor. The inner regions of the disk, closer to the central object, are hotter and emit shorter-wavelength radiation, such as X-rays, while the outer regions are cooler and emit longer-wavelength radiation, such as visible or infrared light. This temperature gradient arises from the balance between energy generation through viscosity and energy loss through radiation. Thin accretion disks are commonly observed in various astrophysical contexts, such as in the vicinity of supermassive black holes at the centers of active galactic nuclei (AGN) or in X-ray binary systems, where a compact object accretes matter from a companion star. Theoretical models based on thin accretion disks have been successful in explaining many observed properties, such as the spectral energy distribution and time variability of the emitted radiation.
\begin{figure}
    \centering
     \includegraphics[width=0.9\textwidth]{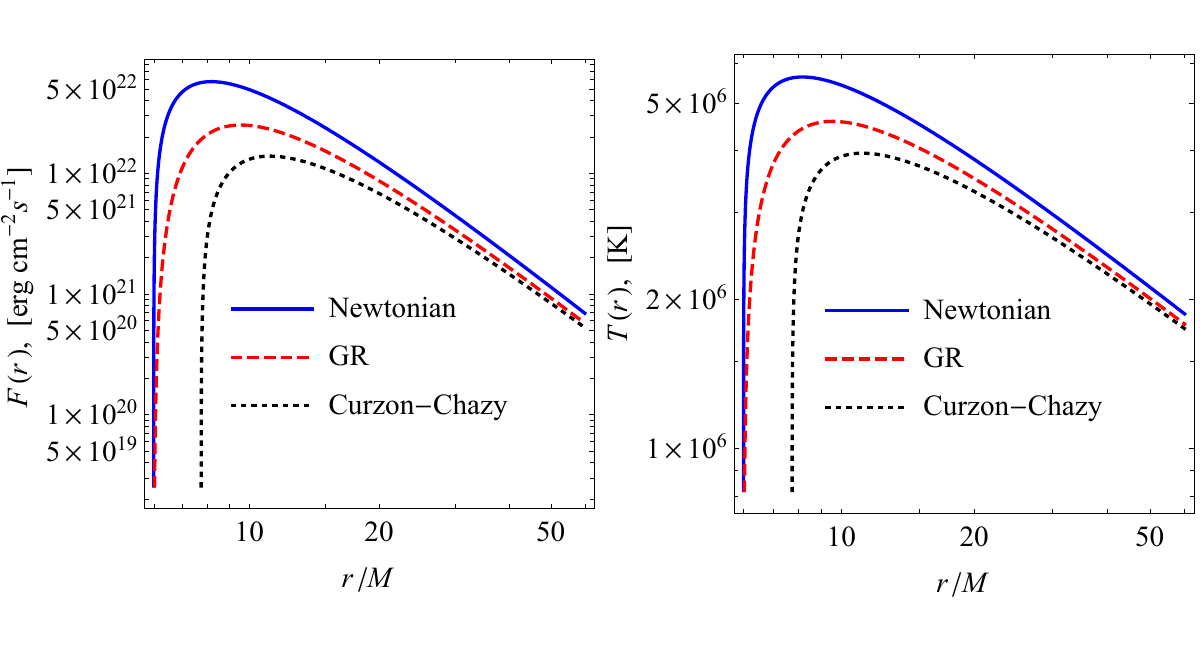}
    \caption{Radial dependence of the the energy flux (left panel) and temperature profile (right panel) in Newtonian and GR frameworks.}
    \label{compare}
\end{figure}

The Novikov-Thorne model, a pivotal framework in the realm of astrophysics, has significantly impacted our understanding of black hole X-ray binaries and active galactic nuclei (AGN) by providing a theoretical foundation for the study of matter accretion onto black holes and the subsequent release of energy in the form of radiation. To model the thin accretion disk within this framework, several key assumptions are made:

(i) Energy loss within the accretion disk occurs as matter spirals inward towards the central object due to friction and gravitational forces. This energy loss leads to the emission of radiation, primarily in the form of X-rays and other high-energy photons.

(ii) The accretion disk is conceptualized as a flattened, rotating structure consisting of gas, dust, and various matter components. As matter migrates inward, it follows nearly circular orbits within the disk, gradually losing angular momentum and moving closer to the black hole. The disk is assumed to be geometrically thin and optically thick, where the radial extension $\Delta r = r_{out} - r_{in}$ is much greater than its thickness ($h << \Delta r$).

(iii) According to the Novikov-Thorne model, an innermost stable circular orbit (ISCO) exists, allowing matter to orbit the black hole or exotic object without rapid infall. The radius of this orbit depends on the mass and angular momentum of the black hole. Matter inside the ISCO swiftly plunges into the central object.

(iv) The innermost part of the accretion disk resides in close proximity to the event horizon of the black hole, the point from which nothing can escape the gravitational pull of the black hole. Radiation emitted by matter near the event horizon experiences significant redshift, making it challenging to detect.

(v) Gas particles in the disk predominantly follow circular Keplerian orbits, rendering them well-described as test particles on circular orbits.

(vi) The torque in the zone near the inner edge of the accretion disk is considered negligible.

(vii) The mass accretion rate of the thin disk is assumed to be relatively constant and should be less than the Eddington mass rate. More precisely, ${\cal\dot M}\simeq (0.03-0.5){\cal\dot M}_{\rm Edd}$, where the Eddington mass rate is defined as $${\cal\dot M}_{\rm Edd}=\frac{4\pi GMm_p}{c\sigma_T}\ .$$ In terms of solar masses, ${\cal\dot M}_{\rm Edd}\simeq 2.33\times 10^{18}\left(M/M_\odot\right) {\rm g\cdot s^{-1}}$.

(viii) The model predicts that the radiation emitted by the accretion disk exhibits a distinctive spectral signature, influenced by factors such as the black hole's mass and spin. This spectrum includes a prominent peak in the X-ray portion of the electromagnetic spectrum.

The Novikov-Thorne model continues to serve as a significant theoretical framework for the examination and comprehension of the characteristics exhibited by thin accretion disks encircling black holes. By employing this model, the radiated energy flux from the disk determined as~\cite{Page1974ApJ,Thorne1974ApJ}.
\begin{align}\label{EnergyFlux}
{\cal F}(r)=-\frac{{\dot {\cal M}}}{4\pi\sqrt{-\tilde{g}}}\frac{\Omega'}{({\cal E}-\Omega {\cal L})^2}\int_{r_+}^r dr ({\cal E}-\Omega {\cal L}){\cal L}' \ ,    
\end{align}
where $\dot{\cal M}$ is the accretion rate of disk, a prime $(')$ denotes the derivative with respect to radial coordinate $r$ and $\tilde{g}=\sqrt{-g_{tt}g_{rr}g_{\phi\phi}}$ is the determinant of the spacetime metric at the equatorial plane. We can find  for the black hole described by rotating Curzon metric through eqs (10) and (14), which is:

\begin{align}
{\cal F}(r)=\frac{\dot{\cal M} \sigma \left(3 r^2-6 r \sigma +4 \sigma ^2\right) }{8 \pi  r^4 (r-2 \sigma )}\left[1-\sqrt{\frac{x_0}{x}}+\frac{2}{x}\left(\tanh^{-1}x-\tan ^{-1}x-\tanh^{-1}x_0+\tan^{-1}x_0\right)\right]e^{\frac{\sigma(\sigma-2r)}{2 r^2}}\ ,
\end{align}
where $x=\sqrt{r/\sigma-1}$. According the Boltzmann law the energy emission rate is proportional to the fourth power of the temperature of the disk i.e. ${\cal F}(r)\sim T^4(r)$. So that the effective temperature of the thin disk is determined as 
\begin{align}\label{temp}
T(r)=\sqrt[4]{\frac{{\cal F}(r)}{\sigma_B}}\ ,    
\end{align}
where $\sigma_B$ is the Stefan-Boltzmann constant. The dependence of the energy flux (left panel) and temperature (right panel) of the accretion disk on the radial distance $ r $ is illustrated in Fig.~\ref{energy flux}. From the figure, it is evident that for larger values of the spin parameter $ a $, both the energy flux and temperature initially increase more rapidly with decreasing $ r $. This is consistent with the fact that higher spin values result in smaller ISCO radii, leading to greater energy release in the inner regions of the disk due to stronger gravitational effects. However, as $ r $ increases, an interesting trend emerges: the energy flux and temperature values for higher spin parameters ($ a = 0.8, 0.998 $) fall below those observed for the non-rotating case ($ a = 0 $). This indicates that, at large radial distances, the energy output in the accretion disk of a non-rotating object surpasses that of a rapidly rotating one. This phenomenon may be attributed to the redistribution of angular momentum and energy dissipation mechanisms within the disk structure in different spacetime geometries. The observed behavior has significant astrophysical implications, particularly for the thermal emission profiles of accretion disks around compact objects. The differences in temperature and energy flux could lead to distinct observational signatures in the spectral energy distributions of black hole accretion disks, potentially allowing future X-ray and optical observations to infer the spin parameter of the central object. High-resolution observations from instruments like NuSTAR, eROSITA, and the Event Horizon Telescope could provide further insights into these variations and help distinguish between different black hole solutions.

In Fig.~\ref{compare}, we compare the energy flux and temperature profiles in three distinct scenarios: the Shakura-Sunyaev model in the Newtonian framework, the Novikov-Thorne approach in the general relativistic case (specifically for the Schwarzschild metric), and the Curzon-Chazy metric. This comparison illustrates that, for the same radial distance, both the energy flux and temperature are generally lower in the Curzon-Chazy metric compared to the other two cases. The Shakura-Sunyaev model, based on a Newtonian approximation, assumes a simpler gravitational potential and leads to higher energy flux and temperature profiles in the inner regions of the disk. The Novikov-Thorne approach, which is based on general relativity and applied to the Schwarzschild metric, accounts for the more complex gravitational effects around a black hole, resulting in a higher energy flux and temperature than in the Curzon-Chazy model. In contrast, the Curzon-Chazy metric, which incorporates additional modifications due to the spacetime's rotation and mass distribution, predicts a lower energy flux and temperature, indicating that the gravitational influence is weaker compared to the Schwarzschild solution. The differences between these models highlight the impact of the underlying spacetime geometry on the thermal and energetic properties of accretion disks. While the Shakura-Sunyaev and Novikov-Thorne models are widely used for describing accretion disks around compact objects, the Curzon-Chazy metric presents a more generalized approach, which could be relevant for studying alternative black hole solutions or objects with non-standard gravitational fields. The lower energy flux and temperature in the Curzon-Chazy metric could offer distinct observational signatures, which might be detectable with current or future X-ray telescopes, offering new insights into the nature of accreting black holes and their surrounding environments.
\begin{figure}
    \centering 
    \includegraphics[width=.45\textwidth]{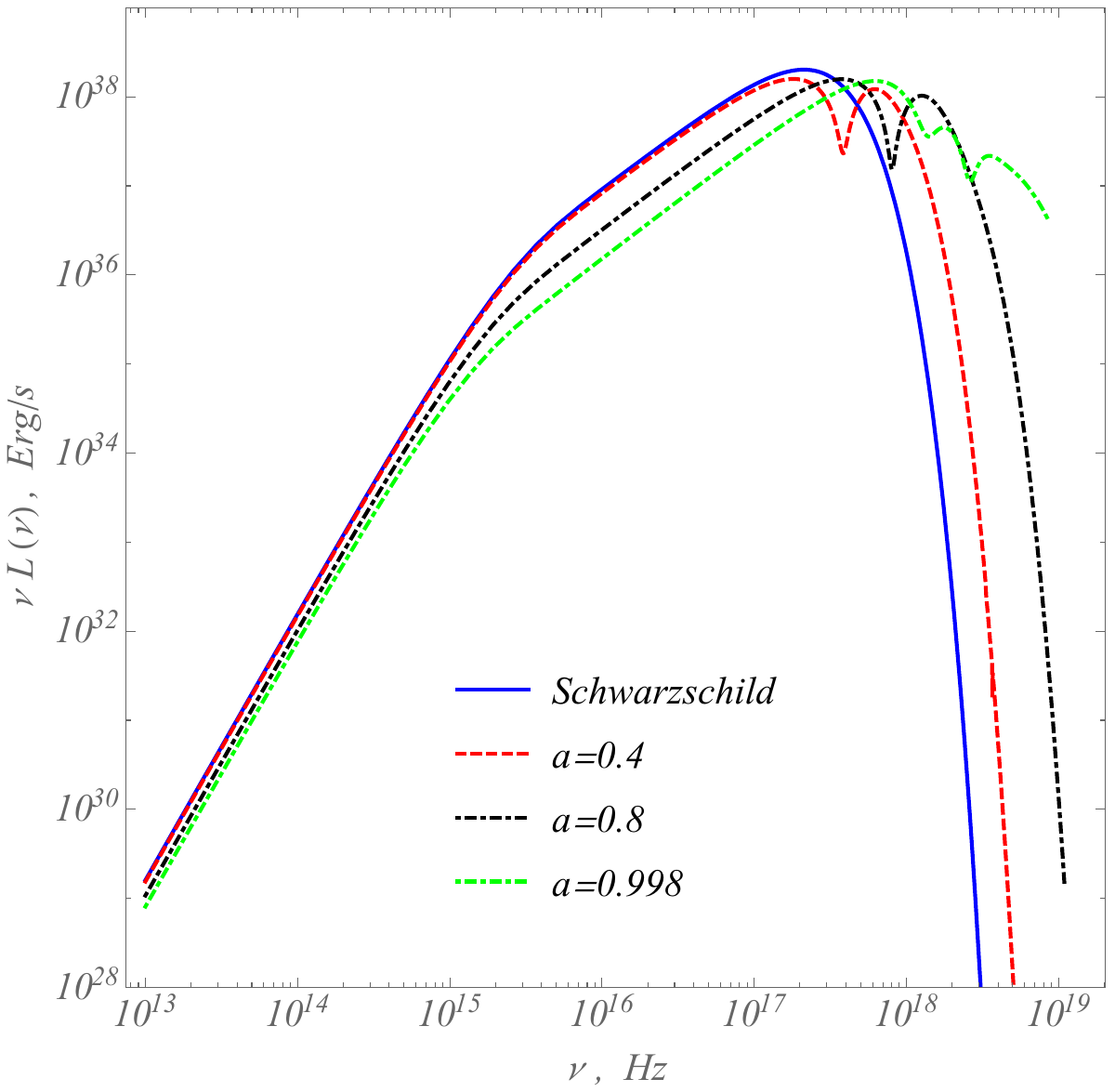}
    \caption{Dependence of the spectral luminosity from frequency of the electromagnetic signal. Upper blue curve represents spectral luminosity in the Schwarzschild metric, while red dashed, black dot-dashed and green curves represent luminosity in Curzon-Chazy metric.}\label{luminosity}
\end{figure}

Using the expression for the temperature profile on the disk, denoted as $T$ and represented by equation \eqref{temp}, and assuming that the disk emits thermal black body radiation, we can derive the spectral luminosity, denoted as $L(\nu)$, of the disk as a function of the frequency of the emitted radiation. This luminosity is given by the following expression:
\begin{align}\label{Luminosity}
L(\nu)=\frac{8\pi h\cos\chi}{c^2}\int_{r_{\rm in}}^{r_{\rm out}}\int_0^{2\pi}\frac{\nu_e^3rdrd\phi}{\exp\left(\frac{h\nu_e}{k_B T}\right)-1}\ ,    
\end{align}
In this equation, $\chi$ represents the inclination angle relative to the symmetry axis, while $r_{\rm in}$ and $r_{\rm out}$ are the locations of the inner and outer edges of the disk. The frequency $\nu$ is measured by a distant observer, and $\nu_e$ is the frequency of photons emitted from the accretion disk. These two frequencies are related to each other through the red-shift factor denoted as $\nu_e=g\nu$, where $g$ can be expressed as follows:

\begin{align}\label{RedShift}
g=&u^t\left(1+\Omega r\sin\phi\sin\chi\right)\ .
\end{align}

Figure \ref{luminosity} provides a visual representation of the total spectral luminosity emitted by a thin accretion disk around a compact object, showing the dependence on both the inclination angle $ \chi $ and the rotation parameter $ a $. As noted, when $ \chi = \pi/2 $, the spectral luminosity vanishes due to the presence of the $ \cos\chi $ factor in equation \eqref{Luminosity}, which essentially causes the observed luminosity to be zero when the observer is aligned with the plane of the disk. In the figure, it is clear that the differences between the spectra for various values of the inclination angle and rotation parameter are most apparent in the high-frequency region, especially within the X-ray band. At lower frequencies, including the infrared (IR), optical, and ultraviolet (UV) bands, the spectral differences are minimal, indicating that these regions of the spectrum are less sensitive to changes in the inclination angle and spin parameter. However, as one moves to higher frequencies, the contrast between different curves becomes more pronounced, suggesting that the rotation parameter $ a $ has a significant impact on the higher-energy radiation emitted by the accretion disk. Furthermore, the frequency at which the peak luminosity occurs shifts depending on the value of the rotation parameter, with higher spin values typically resulting in a shift toward higher frequencies. This dependency implies that it may be possible to infer the spin of a compact object by analyzing the frequency of the peak in the emitted radiation, providing an opportunity for observational tests of black hole geometries. These results highlight the importance of both the inclination angle and the rotation parameter in shaping the emitted spectral luminosity, particularly in the high-frequency regime. Given the sensitivity of X-ray telescopes, such as XMM-Newton, Chandra, and NuSTAR, the observed differences in spectral features could potentially offer a way to distinguish between different black hole models, including the Curzon-Chazy metric, and help refine our understanding of accreting black holes in astrophysical systems.

\section{Summary}\label{Summary}

In this paper, we have tested rotating Curzon-Chazy spacetime through studying test particle motion, weak gravitational lensing effect and properties of the thin accretion disk surrounding spinning central object. Studying of particle motion in Curson-Chazy spacetime has provided valuable insights into the behavior of particles in a unique gravitational environment. The Curson-Chazy spacetime, characterized by its nontrivial geometry and curvature, has challenged our understanding of classical mechanics and general relativity. The curvature of Curson-Chazy spacetime leads to the presence of additional forces acting on particles, altering their trajectories and introducing intriguing effects. These effects can manifest as deflections, precessions, or even unstable orbits, deviating from the predictions of classical mechanics. Moreover, the study of particle motion in Curson-Chazy spacetime has implications for our understanding of fundamental physics. It sheds light on the interplay between gravity and the behavior of particles at extreme conditions, helping us refine and expand our theoretical frameworks. We have studied massive particle motion in rotating Curson-Chazy spacetime. We have presented analytical expression for the characteristic radii, namely, the ISCO and MBO position for massive particle as well as photonsphere for massles particle. It is found that characteristic radii decrease for rotating object and for maximally rotating object they will be zero. 

We have studied gravitational lensing effects near the gravitational object represented by the rotating Curzon-Chazy spacetime. Particularly, we have considered weak lensing and determined a deflection angle of the light-ray in rotating Curzon-Chazy spacetime. It has shown that deflection angle of the light-ray decreases due to the rotating parameter $a/M$ of the gravitational object. Particularly, when spin parameter of the black hole reaches maximal value, the latter does not influence light-ray which is passing nearby and it follows straight line with feeling the external gravitational field. Additionally, our analyses show that in the rotating Curzon-Chazy spacetime, observer overlook three images of single source. However, two of them is similar to each other. So one conclude that absolute value of magnification ratio for two images is the same (lower curve) and they are falling down due to increase value of $\beta$ angle,  while magnification ratio for other image is increasing. On the other hand, total magnification decreases with the increase of both spin parameter $a$ and angle $\beta$. We have also studied gravitational time delay and have shown that it increases due to increase $\beta$ angle. 

Finally, we have conducted an analysis on the characteristics of thin accretion disks surrounding a compact object governed by the rotating Curzor-Chazy spacetime. Given the inherent disorder in particle motion, our investigation has focused exclusively on the equatorial plane in order to ensure integrability. By employing the Novikov-Thorne model, we initially ascertained the rate at which energy is emitted from the surface of the disk. We applied the steady-state Novikov-Thorne model to rotating Curzon-Chazy spacetime and numerically obtained the accretion disk profiles such as the energy flux and temperature distribution for thin accretion disks. The ISCO radius of massive particle has been determined and assumed that it is indicated inner edge of the thin accretion disk. The temperature distribution across the accretion disk was determined using the standard Stefan-Boltzmann law. We have compared to energy flux and temperature of thin accretion disc of spinning black hole in three cases that Newtonian, Shakura-Sunyaev approach. As observations of electromagnetic spectrum become more accurate, study of emission spectra of accretion disks can be a way for testing rotating Curzon-Chazy spacetime. We have shown that if central object have large spin parameter, value of energy flux and temperature is rapidly decrease far from accretion disc.  Subsequently, we computed the total spectral luminosity of this disk for various observer positions and diverse values of the rotating Curzon-Chazy parameter. Our findings indicate that the deformation of spacetime has the potential to influence the emission in the high-frequency range originating from the thin accretion disk. 

%\section*{Acknowledgement}The research is supported by the Grants F-FA-2021-510 from the Uzbekistan Ministry for Innovative Development.
%\section*{Data availability}{This research has no associated experimental and observational data.} 
%\section*{Conflicts of interest}{The authors declare no conflict of interest.} 

\bibliographystyle{model}
\bibliography{Ref}
\end{document}